# Upconverting nanodots of NaYF$_4$:Yb$^{3+}$Er$^{3+}$; Synthesis, characterization and UV-Visible luminescence study through Ti: sapphire 140-femtosecond laser–pulses

*Monami Das Modak$^a$, Ganesh Damarla$^b$, K Santhosh Kumar$^a$, Somedutta Maity$^a$, Anil K Chaudhury$^b$ and Pradip Paik$^{a,c*}$*

$^{a.}$School of Engineering Sciences and Technology, University of Hyderabad, Hyderabad, Telangana, PIN: 500046

$^{b.}$ Advanced Center of Research in High Energy Materials, University of Hyderabad, Hyderabad, Telangana, PIN: 500046 Address here

$^{c.}$School of Biomedical Engineering, Indian Institute of Technology (IIT)-BHU, Varanasi, UP, PIN 221005

**ABSTRACT**: In this work, dot-sized upconversion nanocrystals (UCN-dots) with diameter c.a. 3.4±0.15 nm have been synthesized. These UCN-dots exhibit visible emission (at λ = 497, 527 and 545 nm) under the excitation with 980 nm CW-NIR laser. Further, these UCN-dots exhibit high energy upconversion emission (UV region, λ = 206 to 231 nm) with Ti-Sapphire Femtosecond laser of 140-femtoseconds duration at 80 MHz repetition rate at different excitation, which has never been reported. This is interesting to report that the generation of high energy UV-Vis emission and their shifting from λ = 206 to 231 nm for the UCN-dots by tuning the excitation wavelength ranging from λ = 950 nm to 980 nm irradiated from Ti: sapphire Femtosecond laser observed. We have demonstrated the generation of high energy upconversions with change in energy band gaps as well as number of absorbed photons per photon emitted under the Femtosecond-laser excitation power. Additionally, we report the photo luminescence of UCN-dots in visible range with 450 nm excitation wavelength exhibiting blue and red emission (visible to visible). The generation of high energy up-conversion in UV-Vis region could be useful for designing optoelectronic and biomedical devices for therapeutic application.

## INTRODUCTION

Upconversion phenomenon is a two photon-(one from sensitizer ion and other one from activator ion) process followed by Energy Transfer (ET) which contributes higher photon energy in ultraviolet (UV) and visible range under the excitation with Infrared (IR) or near infrared (NIR) radiation. The transmutation of IR/NIR photons to UV/visible-photons in upconversion nanoparticles (UCNPs) make them unique for applications in electronics and biomedical. Though there are a few lanthanide ions manifested upconversion but substantially $Yb^{3+}$, $Er^{3+}$, $Tm^{3+}$ trivalent lanthanide ions have been used to synthesize upconversion nanocrystals due to their efficient upconversionemission.[1–3] Till date, $NaYF_4$: $Yb^{+3}$, $Er^{+3}(Tm^{+3})$ has been proven to be one of the most efficient UCNPs[3–5] due to its highest photon upconversion efficiency. It is well known that smallest-particles have always been demanded and they attract nano medicine for therapeutic applications[6,7] including bio-labelling,[8] in-vivo imaging,[9] bio-conjugation,[10] long term cell tracing,[11,12] and bio-detection.[13] Owing to strong biological relevance UCNPs are used for both *in vitro* and *in vivo* applications along with molecular bio-imaging and for targeted cancer therapy.[14–17] Therefore, it would be a superior idea to combine both the approaches. It is also reported that UCNPs can be used for making solar cells, photovoltaic and plasmonic devices and to increase the efficiency of display devices.[18–23]

Here in this work UCN-dots have been synthesized through a one-pot chemical-synthesis approach. These UCN-dots are having diameter below 4 nm. All the other works reported are on the synthesis of UCNPs of size above 10 nm (dia.) and are mostly with hexagonal crystalline phases.[1–3,5,24] However, UCN-dots reported in this work have been synthesized within a very short period of time. Short reaction period at moderately high temperature avoids the production of large NaF crystal-matrix and it could also avoid the production of other fluorinated oxygen and carbon species during synthesis. The second objective of this work is to study the upconversion properties of UCN-dots with femtosecond laser source (Ti-Sapphire Femtosecond laser of 140-femtoseconds duration at 80 MHz repetition rate).

## EXPERIMENTAL:

## MATERIALS AND METHODS

### Materials

$YCl_3:6H_2O$; $YbCl_3:6H_2O$ and $ErCl_3:6H_2O$ precursors (Sigma–Aldrich and with 99% purity), 1-Octadecene (90%), Oleic Acid (65%) NaOH (97.0%), $NH_4F$ (95 %) were purchased from sigma Aldrich, Qualigens, SDFCL and Kemphasol, respectively were used for the synthesis of UCN-dots.

### Synthesis of UCN-Dots.

The detail synthesis procedure of UCN-dots has been filed for Indian patent (Ref. TEMP/E-1/21071/2017CHE, dt.: 14/06/2017). In brief of the synthesis procedure: the specific amount of precursor materials ($Ycl_3:6H_2O$; $YbCl_3:6H_2O$ and $ErCl_3:6H_2O$) were dried from moisture and poured in 1-Octadecene and Oleic-acid followed by heating in inert environment. Then in cold condition a mixture of 4:1 mol of NaOH and NH4F (dissolved in MeOH) was added. The resulted solution was heated in inert gas environment up to 300°C for few mins and then it was cooled. Finally the UCN-Dots were collected via high speed centrifugation (RPM 14000) and were preserved in cyclohexane.

**Characterizations.**

The particle size and morphology of UCN-dots were characterized through transmission electron microscopy and high resolution transmission electron microscopy (HRTEM) (model FEI TecnaiG2-TWIN 200 KV), elemental analysis was performed using Energy Dispersive X-ray analysis (EDXA), X-Ray diffraction pattern (XRD with CoKα-radiation) was used to determine the crystal structure. Raman Spectroscopy (WiTec, alpha 300) was used to find out the solid state lattice vibration and to find out the defects, Fluorescence Spectrophotometer (HITACHI, F-4600) provided with an external near infrared (NIR) laser source 980nm to measure the upconversion fluorescence (UCF) spectra in visible region. Ti Sapphire tunable oscillator laser with Femtosecond pulses were used to study the upconversion in UV and Visible region. The arrangement of Femtosecond experimental set-up has been shown in Figure 6.

**RESULTS**

UCN-dots were synthesized as discussed in the experimental section and characterized them through the suitable methods. Figure 1(a-c) show the TEM micrographs of the synthesized UCN-dots and their particle size distribution, respectively. UCN-dots (NaYF$_4$:Yb$^{3+}$; Er$^{3+}$) possess uniform size distribution confining size below 4 nm in diameter and they are monodispersed in nature. These UCN-dots are of average size ~3.4±0.15 nm in diameter (Figure S1). These UCN-dots are stable up to one year due to their high surface zeta potential of -36.39 mV (Figure S2) Figure 1(c-d) shows high resolution TEM (HRTEM) micrograph exhibited clearly the lattice fringes and confirming the crystallinity of the UCN-dots. Distance between two adjacent fringes has been calculated to be 3.1 Å and is matching well with the results reported for larger sized UCN nanoparticles.[1] Figure 1(e) shows SAED pattern (ring pattern) corresponding to the (111), (220), (222), (331)and (420) reflection planes for pure cubic crystal structure, which further have been confirmed through the XRD and Raman spectroscopy and explained in the subsequent sections.

Crystal structure of solid powder UCN-dots has been confirmed through XRD study Figure 2. XRD pattern confirms the well-defined diffraction peaks at 2θ = 31.8°, 45.5° and 66.3° correspond to the reflection planes (111), (220) and (222), respectively and represented for the cubic (FCC) crystalline phase of the UCN-dots. Crystallite size of UCN-dots has been calculated using Debye-Scherer formula[2,25] and the average crystallite size has been found to be ~3.4±0.15 nm which is matching well with the results of the particle size obtained from TEM (Figure 1).

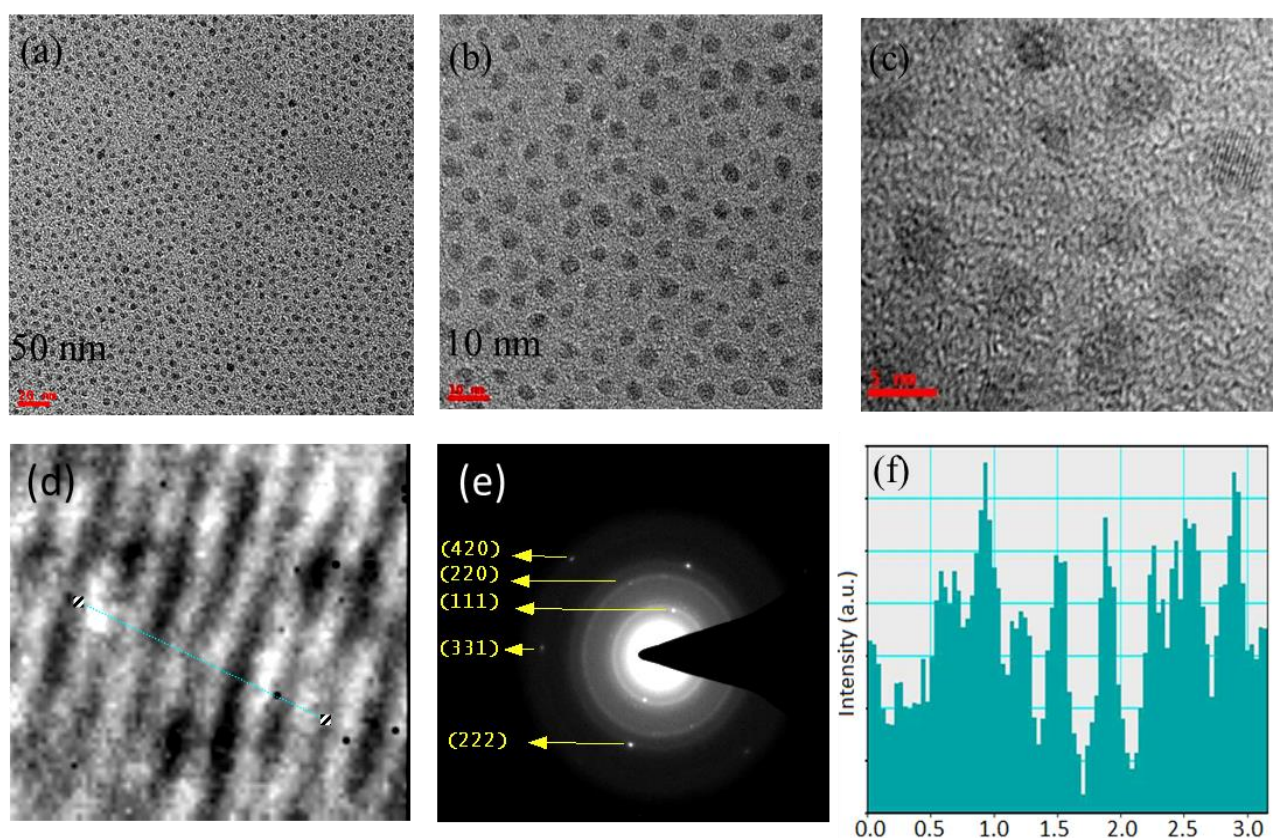

Figure 1. Shows TEM results. Fig. (a), (b) and (c) for TEM micrograph of UCN-dot from low to high magnification.  (c) HRTEM shows the crystalline phases of the particles. (d) HRTEM image and fringe

5 | P a g e

width. (e) Selected area electron diffraction pattern (SAED) obtained from TEM and (f) profile of the fringes obtained from figure (d).

The Lattice parameters for cubic crystalline UCN-dots have been evaluated to be a= 5.51 Å, b= 5.31 Å and c=5.32 Å which is matching well for the cubic phases of UCN nanoparticles reported.[1,26] The elemental composition of the UCN-dot also confirmed from the EDAX analysis (Figure S3 and Table S1). From XRD (Figure 2) and from the EDAX (Figure S3) results the number of unit cell present in a single UCN-dot has also been calculated and found to be: 143. It can be noted that the number of different atoms present in a unit cell of UCN-dot are: 2, 8, 1 and 1 for Na, F, Yb and Er, respectively as a fraction of $Y^{3+}$ ions are substituted by rare earth ions ($Yb^{3+}$ and $Er^{3+}$).[27]

Figure 3 shows the Raman spectrum of the UCN-dots in different band regions. Figure 3a, shows the entire spectra where most of the bands appeared between 100-1000 $cm^{-1}$ representing the appearance for the cubic phases of UCN-dots ($\alpha$-$NaYF_4$:$Yb^{3+}$;$Er^{3+}$) whereas, band region appeared between 1000 $cm^{-1}$ -3500 $cm^{-1}$ confirms for the defects of UCN-dots ($\alpha$-$NaYF_4$:$Yb^{3+}$;$Er^{3+}$) and the presence of oleic acids (capping agent).

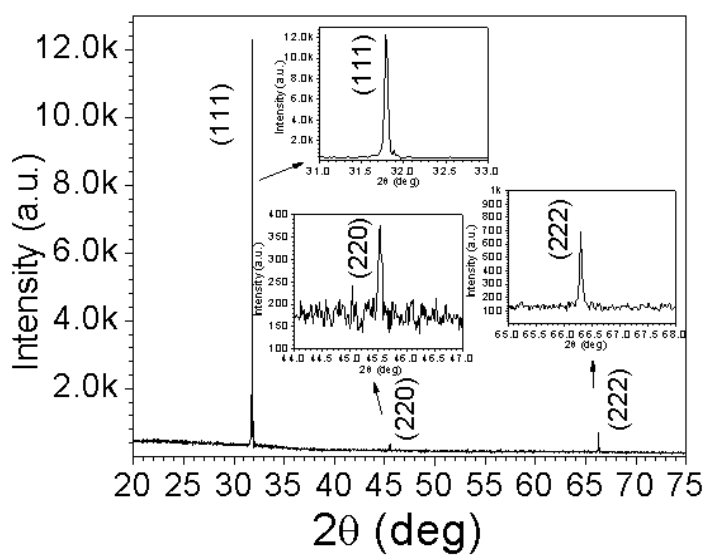

Figure 2. XRD of UCN-dots represents the crystalline phase (cubic, FCC) of UCN-dot. In-sets are showing enlarge peaks for (111), (220) and (222) planes

Weak Raman bands with phonon frequency below 1000 cm$^{-1}$ confirm for the presence of vibration modes (ν) of α-NaYF$_4$ or β-NaYF$_4$ crystals.[28] Hence, below 1000 cm$^{-1}$ phonon vibration bands appeared at 265 cm$^{-1}$, 326 cm$^{-1}$, 480 cm$^{-1}$, 567 cm$^{-1}$, 636 cm$^{-1}$ and 752 cm$^{-1}$ (Figure 3b) further satisfied the presence of major α-phase along with relatively less extent of β-phases in UCN-dots.[28,29]

However, presence of weak β-phases is identified with band positions at 480 cm$^{-1}$ and 636 cm$^{-1}$. Other bands are appeared at 326 cm$^{-1}$ and 567 cm$^{-1}$ which can be ascribed as distinct vibrations originated from synthesized UCN-dots with α-phase. The high frequency phonon energy modes appeared at 1085 cm$^{-1}$ and 1305 cm$^{-1}$ correspond to the presence of C-F and –CH$_2$ groups. The bands appeared at 1449 cm$^{-1}$, 1639 cm$^{-1}$, 1803 cm$^{-1}$, 2173 cm$^{-1}$ and 2701 cm$^{-1}$ due to the presence of CH$_2$, -CH$_3$ groups, -C=C- bond, -C=O bond, -C≡C-and–CHO, respectively of oleic acid (capping agent) (Figure 3c and Figure 3d). Two weak intensity bands appeared at 1998 cm$^{-1}$ and 2428 cm$^{-1}$ due to the possible defects present in UCN-dots (Figure 3c).

However, bands for β-phase of NaYF$_4$:Yb$^{3+}$;Er$^{3+}$ are found very weak and only bands for prominent α-NaYF$_4$:Yb$^{3+}$;Er$^{3+}$ cubic phases[28–31] are present in UCN-dots. Further, the weak bands appeared at 687-703 cm$^{-1}$ and 260-279 cm$^{-1}$ along with the broad band appeared in between 750-1750 cm$^{-1}$ confirmed for the FCC structure of α-NaYF$_4$:Yb$^{3+}$;Er$^{3+}$ (phonon vibrations in between 1384-1416 cm$^{-1}$ and 932-1041 cm$^{-1}$ with moderate intensity).[29,31] Thus for Raman it is confirmed that the UCN-dots prepared in this work is with cubic crystalline phase which) complied with the results obtained from the TEM (Figure 1e and XRD (Figure 2).

The possible reasons for the appearance of higher frequency bands in Raman spectra in UCN-dot particles are: (i) the dimensions of particles are very small due to which a higher number of capping agents are being absorbed on the surface, as a result high energy C-H or C-C vibrations can be caused of getting higher frequency Raman bands, (ii) presence of functional groups (O=C-O$^-$ and OH$^-$) of oleic acid attached on the surface of the particles and (iii) finally, a higher percentage of capping agents and their functional groups on the UCN-dots can act as surface active agents which can initiate a compressive stress leading to closely packed surface atoms, and as a result of vibration, Raman bands appeared at extended band position.

Figure 4a shows the photo luminescence (PL) spectra of UCN-dots in aqueous medium obtained with an excitation wavelength of λ= 450 nm and their corresponding energy transitions within electronic states has been shown in Figure 4b. Photo luminescence bands are observed mostly in visible region. The PL spectra have been acquired within 380 nm-700 nm region.

Figure 3. Raman Spectra for UCN-dots. (a) Spectra for the full region (250 to 3750 cm$^{-1}$) (b) band region

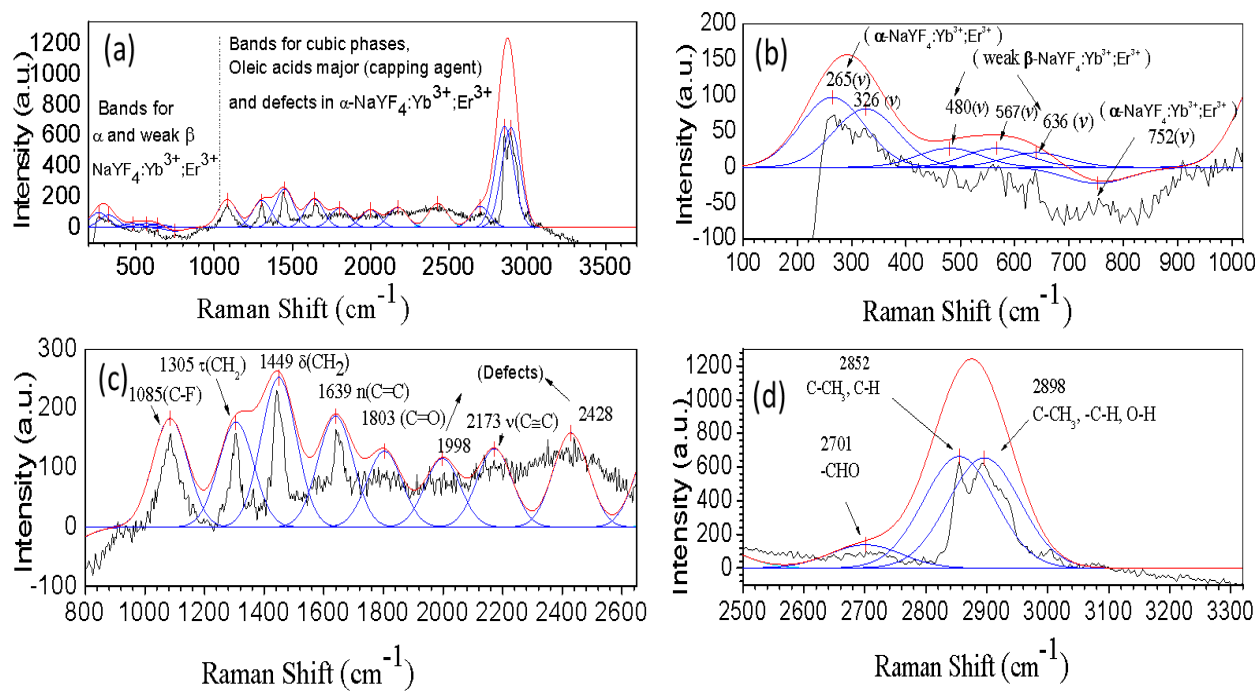

100 cm$^{-1}$ to 1000 cm$^{-1}$ representing the major α-phase of NaYF$_4$:Yb$^{3+}$;Er$^{3+}$ dots, (c) 800 cm$^{-1}$ to 2600 cm$^{-1}$ and (d) 2500 cm$^{-1}$ to 3300 cm$^{-1}$, exhibiting the presence of major capping agents in cubic phases

The obtained major bands are centred at 407 nm, 430 nm, 462 nm, 520 nm, 548 nm, 570 nm, 607 nm, 640 nm following two direct energy transfers from $Yb^{3+}$ ion to $Er^{3+}$ ion (Figure 4a and 4b). NIR-NIR upconversion photo-luminescence was reported for $NaYF_4$:$Yb^{3+}$, $Er^{3+}$ UCNPs elsewhere.[32–34] However, excitingly in the present study the visible to visible upconversion photo luminescence is also observed for UNC-dots while a 450 nm excitation source was used. The highest PL emission bands are observed at 462 nm ($E_g$=2.68 eV) and at 640 nm ($E_g$=1.94 eV) a strong intense emission band is observed. Additionally, few more PL emission bands appeared relatively with weak intensity.

From Figure 4a and 4b the emission band appeared at 520 nm ($E_g$=2.38 eV) and can be attributed to the energy transition $2H_{11/2} \rightarrow 4I_{15/2}$ (Figure 4b), and other two bands appeared at 548 nm ($E_g$=2.26 eV), 570nm ($E_g$ =2.18 eV) correspond to the $4s_{3/2} \rightarrow 4I_{15/2}$ transition and these three are responsible for the green emission. Whereas red emission band appeared at 607 nm ($E_g$= 2.04 eV) and 640 nm ($E_g$ = 1.94 eV) which is configured for the $4F_{9/2} \rightarrow 4I_{15/2}$ transition (Figure 4b). Emission- bands appeared at ~407 nm ($E_g$ = 3.05eV), 430 nm ($E_g$=2.88 eV), 462 nm ($E_g$=2.68 eV) satisfied the energy transition $2H_{9/2} \rightarrow 4I_{15/2}$.[35] Interestingly, one band with moderate intensity appeared (in near-UV region) at 380 nm ($E_g$ = 3.26 eV) corresponds to the energy transfer $4G_{11/2} \rightarrow -4I_{15/2}$ following a transitional energy transfer (indirect energy transfer) from $2H_{9/2} \rightarrow 4G_{11/2}$ (Figure 4b). However, the corresponding emission bands observed from PL has been compared with fluorescent spectra (FL) obtained with the interaction of CW-980 nm laser and discussed in the subsequent sections.

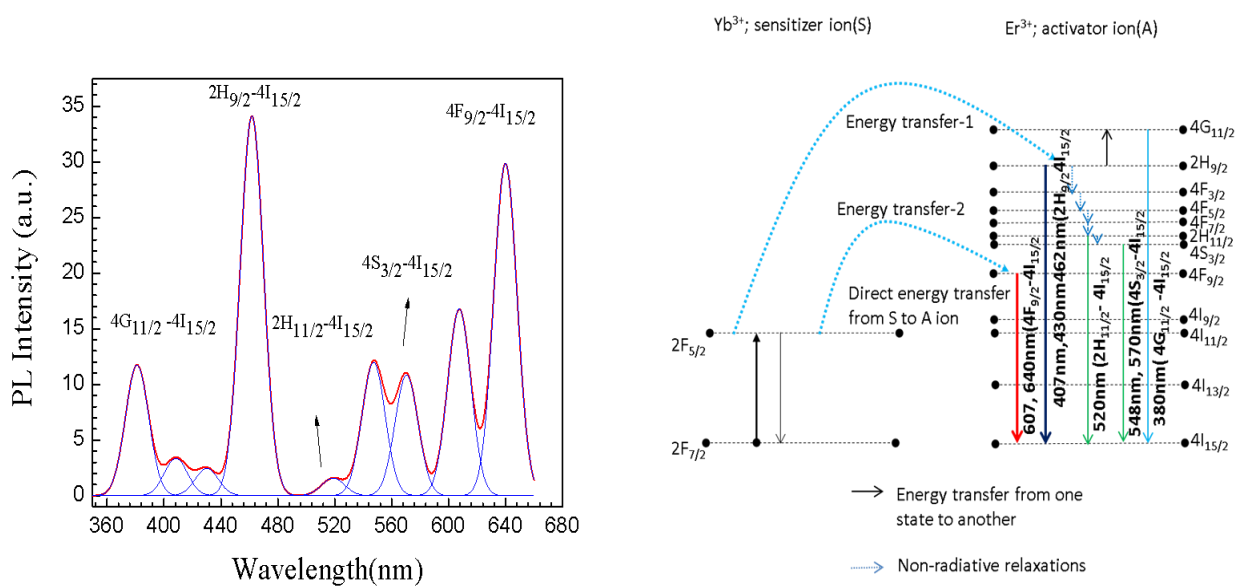

Figure 4. (a) PL spectra of the UCN-dots of size 3.5 nm. The spectrum was obtained with excitation wavelength of 450nm. The peak positions represent the different transitions of the photons at the visible region (b) energy transitions within electronic states of PL spectra obtained with excitation wavelength of 450nm

**Upconversion mechanism (CW-980 nm laser)**. Up-converted materials qualify the inclusion of two or more photons resulting in the emission of higher energy photons through several energy transfers (radiative and non-radiative) either between two ions (activator ion and sensitizer ion) or within the energy states of an activator ion itself in the visible-spectrum range which has been shown in Figure 5a for UCN-dots. For UCN-dots (NaYF$_4$:Yb$^{3+}$; Er$^{3+}$) the Yb$^{3+}$ ion acts as a sensitizer and Er$^{3+}$ acts as an activator.[1,35,36]

In fluorescence spectrum (FL) (see Figure 5a), the different emissive bands denote the intensity of the emitted higher energy photons at different wave lengths depending on the transfers of electrons from different excited states to the ground state or first excited state. Fluorescence emission spectra of colloidal solutions of UCN-Dots display well separated emissive peaks at room temperature. Only four significant emission bands are identified under the CW-laser excitation source (980 nm).

The energy transfer mechanism has been shown in Figure 5b. The green emissions in between 513 nm to 533 (527 nm, $E_g$ = 2.35 eV) nm and 533 nm- 569 nm (545 nm; $E_g$ = 2.28 eV) can be assigned to the $2H_{11/2} \rightarrow 4I_{15/2}$ and $4S_{3/2} \rightarrow 4I_{15/2}$ transitions, respectively. A dominant red emission is observed between 630 nm and 680 nm (663 nm; $E_g$ = 1.87 eV) and can be assigned to a transition $4F_{9/2} \rightarrow 4I_{15/2}$. A feeble

NIR emission between 795 nm and 883 nm (838 nm; $E_g$ = 1.48 eV) is appeared due to the $4S_{3/2} \rightarrow 4I_{13/2}$ transition of the photons. A highly intense emission band is appeared in between 485 nm to 506 nm (496 nm; $E_g$ = 2.5 eV) can be assigned to the transition $4F_{5/2} \rightarrow 4I_{15/2}$, which occurs through a direct energy transfer from $2F_{5/2}(Yb^{3+}) \rightarrow 4G_{11/2}(Er^{3+})$ by continuous 980 nm input excitation laser source. From the higher excited state $Er^{3+}$ ion can relaxed to the next lower states following the path $2H_{9/2} \rightarrow 4F_{3/2} \rightarrow 4F_{5/2}$ (due to close proximity of these intermediate levels) through several non-radiative multiphonon relaxations which do not emit any photons. The emission band appeared in between 485 nm and 506 nm can neither be fallen exactly under green emission band nor to blue emission due to a divergent wavelength range. Therefore, we cannot assign them directly to the transitions $2H_{11/2}/4S_{3/2} \rightarrow 4I_{15/2}$ or $2H_{9/2} \rightarrow 4I_{15/2}$. An intermediate or reservoir state ($4F_{5/2}$, $4F_{3/2}$) should be responsible for such emissions which can be assigned to the transition $4F_{5/2} \rightarrow 4I_{15/2}$ (an immediate cross relaxation occurs between $4F_{3/2}$ and $4F_{5/2}$ levels due to their close proximity)  and follows  the energy transfer path:

$2F_{7/2}(Yb^{3+}) \rightarrow 2F_{5/2}(Yb^{3+}) \rightarrow 4G_{11/2}(Er^{3+}) \rightarrow 2H_{9/2}(Er^{3+}) \rightarrow 4F_{3/2}(Er^{3+}) \rightarrow 4F_{5/2}(Er^{3+}) \rightarrow 4I_{15/2}(Er^{3+})$ which includes three non-radiative relaxations such as (i) $4G_{11/2}(Er^{3+}) \rightarrow 2H_{9/2}(Er^{3+})$, (ii) $2H_{9/2}(Er^{3+}) \rightarrow 4F_{3/2}(Er^{3+})$ and (iii) $4F_{3/2}(Er^{3+}) \rightarrow 4F_{5/2}(Er^{3+})$ (Figure 5b).

Nearly 50% absorbed NIR photons are upconverted from the UCN-dots resulting in visible emission spectrum. As the synthesized particles are of dot- sized particles (size~ 3.5 nm), therefore, number of non-radiative relaxations increase by the solvent molecules and an overall quantum yield (Q.Y.) decreases.

The energy level diagrams of UCN-dots and their up-conversion mechanisms following CW-980 nm laser diode excitation is shown in Figure 5b, where the energy transitions occurred through several radiative and non-radiative emission pathways. As reported earlier, for $NaYF_4:Yb^{3+};Er^{3+}$ nanocrystals maximum two to three emission bands were obtained from their entire fluorescence spectrum region.[1,2,35] However, for our UCN-Dots four significant emission bands have been appeared (Figure 5a). The intensity ratios of green emissions from 513 nm-533 nm and 533 nm- 569 nm to red emission from 630 nm-680 nm yielded much higher values (such as 1.75 and 2.67, respectively) compared to the results reported for UCN nanoparticles with larger sizes.[1,2,35] In previous reported a size dependent relation of GRR (intensity ratio of green to red emission) in $Yb^{3+}$; $Er^{3+}$co-doped $NaYF_4$ nanocrystals and a graphical plot of GRR as a function of particle sizes has also been observed, where the green emission became more pronounced with decreasing particle size.

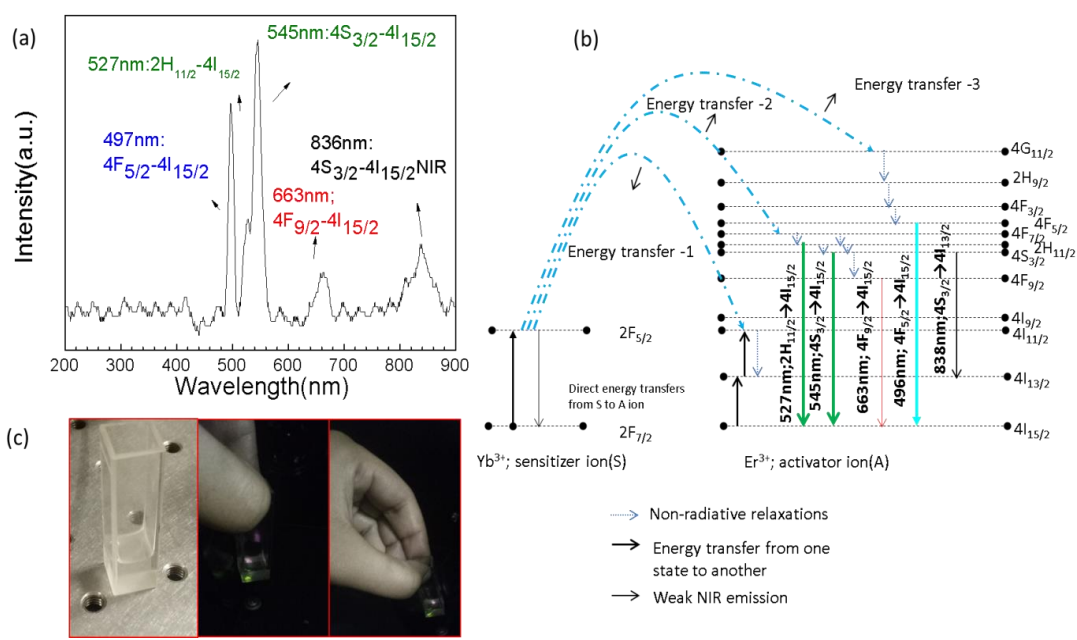

Figure 5. (a) Fluorescence Spectra of UCN-Dot-colloidal solution in cyclohexane with a 980 nm continuous wave (CW) NIR excitation (Power Density 1000 mW/ cm2) exhibiting four significant emission bands at room temperature (b) The relevant energy diagram of dot-sized upconvrting nanocrystals and upconversion processes succeeding 980 nm laser-diode excitation. Further it predicted energy transfers (with three direct energy transfers) with their corresponding pathways (c) Three images are with colloidal solutions and showing the visible emissions under 980nm laser CW.

The GRR value of the present UCN-dots appeared to be very high relative to the UCN nanoparticles with larger size.[35] The GRR value of the present UCN-dots appeared to be high compared to the larger sized UCN particles.[35] This phenomenon can be explained as: the dot-sized UCN particles are responsible for an increased number of non-radiative relaxations as the dot-sized nanoparticles are having higher specific surface area to volume ratio. As a result an increased number of doped lanthanide ions are closer to their periphery (surface) leading to more number of non-radiative relaxations (as shown in Figure 5b with energy–diagram) between their different energy levels. As a result, the overall fluorescence intensity decreases. Please see the image (Figure 5c) for the visible emissions under 980 nm laser-diode excitation source.

**Up-conversion due to Femtosecond laser (FS-Laser) irradiation and High-Power band generation in UV-Vis region.**

A Ti Sapphire tuneable oscillator laser with pulse duration ~140 femtoseconds at a repetition rate of 80 MHz (coherent chameleon ultra-II made) was used as a pumping source for sample excitation. Luminescence signal was recorded using non-gated Spectrometer (Ocean Optics, MAYA 2000). The schematic of experimental setup is shown in Figure 6. Luminescent results exhibited additional luminescence band appeared at UV-Vis region (205 nm-231 nm), other than the appearance of a number of bands in the visible region (Figure 7).

The generation of emission bands in the UV region in interaction with femtosecond pulses is due to the non-linear effects introduced by the transient second order non-linearity by femtosecond pulses. Further it is noticed that the luminescent bands at UV-Vis region shows shifting of band with respect to the incident pump wavelengths, which is further clearly confirmed in support of the process of high energy upconversion for UCN-dots and this phenomenon is reported here for the first time.

Aktsipetrov et .al reported the magnetization induced harmonic generation and THz in Bi:YIG- a magneto photonics crystals.[37] If these samples are subjected to intense beam of laser then higher order optical harmonics are generated. The first shows the linear response due to transmission while the strongest non-linear response is attributed to the higher order non-linearity. However, all these processes are applicable in case of non-Centro symmetry materials. Most of the materials possess odd order non-linear susceptibility due to the broken crystal tractions, defects external field effects etc. In non-Centro symmetry materials the response is from the structure, however, in case of nanomaterials, structure relation response become more important and size reduction helps to generate the efficient higher order harmonics which is not clear at present study. The non-linear magneto optical effect could be the reason of getting higher order non-linear response from the UCN-dots and as a result upconversion bands appeared at UV-region.

In Figure 7(a-b), different emission bands of colloidal UCN-dots have been observed under the excitations of four different incident pump wavelengths such as $\lambda$ = 950 nm, 960 nm, 970 nm, and 980 nm in femtosecond laser (fs-laser) setup (Figure6). We have observed emissive bands in both the ultraviolet (UV) and visible regions.

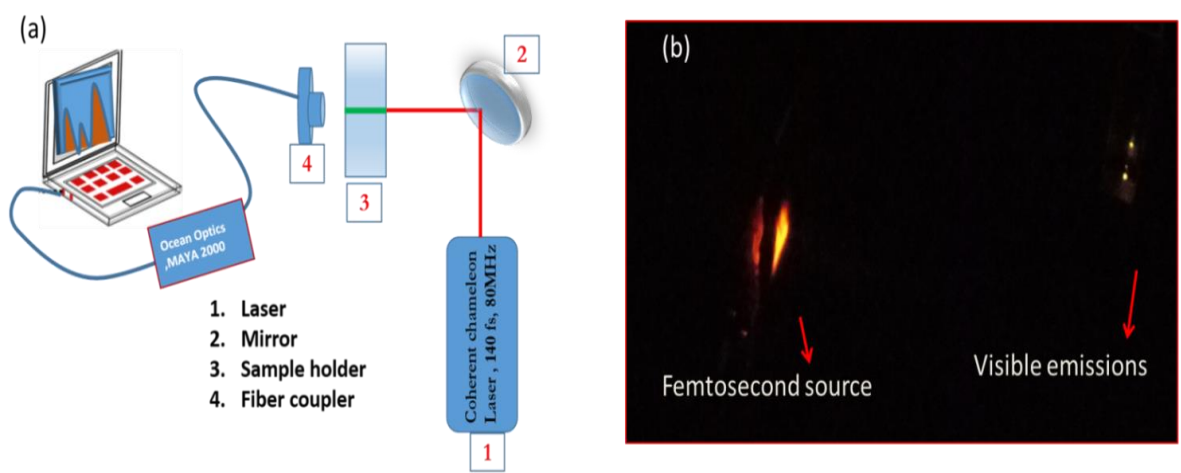

Figure 6. (a) Schematic diagram of a complete experimental set up for Femtosecond laser study (b) Visible emissions (right) with Femtosecond laser source (left) during experiment

In UV region (1st series of the observed emissive bands) (Figure 7a), at λ = 950 nm, 960 nm, 970 nm and 980 nm fs-excitation wavelengths, the emission bands appeared at 206 nm, 212 nm, 227 nm and 231 nm, respectively. Thus, a clear shifting of bands from 206 nm to 231 nm is observed as the excitation wavelength changes from λ = 950 nm to 980 nm. Further, a number of emissive bands also appeared in visible region (Figure 7b). These appeared emission bands in visible region are with lesser intensity compared to the band appeared in the UV region, and their intensities are significant. At four different excitation wavelengths the 2nd series of emission bands (G1) appeared in the wavelength range between 506-535 nm with highest band position at 521 nm, 522 nm, 522 nm and 524 nm, along with a 3rd series of bands (represented as G2) appeared between 536 nm-562 nm with maximum intensity band positions at 540 nm, 540 nm, 541 nm and 542 nm, for the 950 nm, 960 nm, 970 nm and 980 nm excitations, respectively. Thus, at visible region maximum of 2 nm band shifting has been observed. Further, a 4th series of emissive bands (in visible region) appeared in between 643 nm-677 nm having highest intensity band positions at 662 nm, 662 nm, 662 nm and 663 nm under the fs-laser irradiation of 950, 960, 970 and 980 nm, respectively which are not comprised any significant shifting with change in the irradiation wave length (Figure 7b).

Thus, the main interesting attraction is the emission at the UV region once the UCN-dot excited under the Fs-laser and their remarkable red shifting once they are excited from 950 nm to 980 nm Fs-laser irradiation (Figure 7a). It is very exciting that there is no shifting of band positions observed in the visible region for G1, G2 or R bands (Figure 7b). Moreover, band appeared in the UV region exhibited noticeable red shifting as λ increases from 950nm to 980nm irradiation (Figure 7a). This exciting phenomenon can be explained by calculating the number of photons absorbed under the Fs-laser excitation, which is further strongly dependent on: (i) Fs-laser power and number of excited photons and (ii) on the shifting of band gap (Eg= band gap energy).

To measure the Fs-laser power dependent luminescence and the number of absorbed photons taken part in the process, the following power law equation has been used

$$I_{UC} \sim I_{fs} \qquad (1)$$

Where $I_{UC}$ is the luminescence intensity, which is proportional to the 'n$^{th}$' of excitation power $I_{fs}$ 'n' is the number of absorbed photons per photon emitted under the Fs-laser excitation power $I_{fs}$ and this has been calculated from the slopes of $\log(I_{UC})\ vesrsus\ \log(I_{fs})$. Similar power law concept has been used to calculate the NIR laser power dependent luminescence using NIR elsewhere.[38–42]

It can be noticed from Fig 9 (a-d) that for the G1 and G2 luminescence bands we need to transfer a minimum pair of photons (n = 2) from one sensitizer ion i.e., from Yb$^{3+}$ (donor ions,)

16 | P a g e

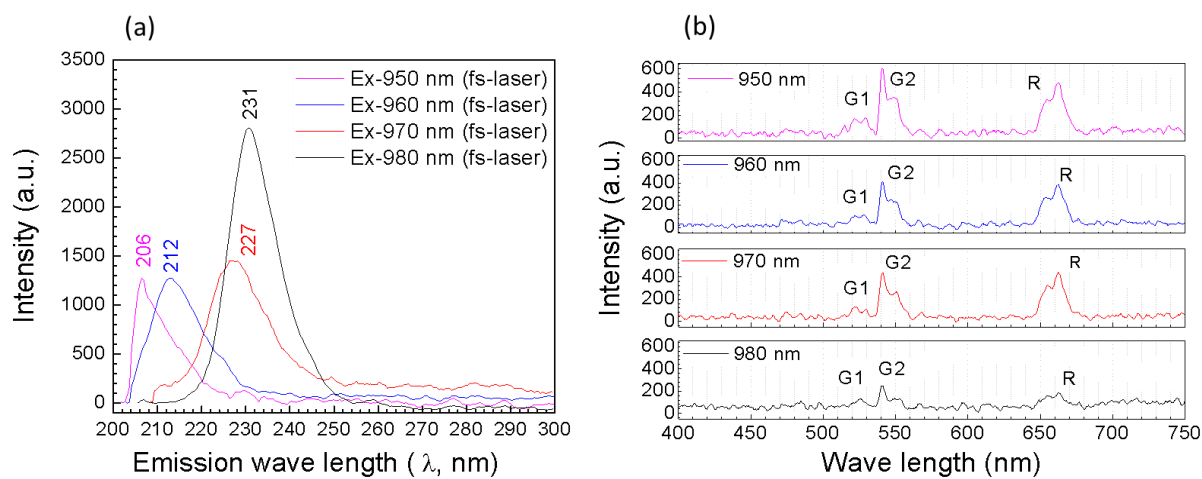

Figure 7. (a) Up-conversion Luminescence spectra in the UV-region (on excitation with fs-laser), exhibited red shifting with increasing excitation wavelengths (b) Up-conversion Luminescence spectra at the visible region for different Fs-laser excitations

to one activator ion i.e., to $Er^{3+}$(acceptor ions) (1: 1 molar ratio) under Fs-laser (NIR) excitation.[39–41] And the values obtained to be n (G1) = 3.6 and n(G2) = 1.6 respectively (Figure 9, $\log(I_{UC})$ vsrsus $\log(I_{fs})$ plots).

Energy transfer 1: $2F_{7/2}(Yb^{3+}) + 4F_{7/2}(Er^{3+}) \rightarrow 2F_{5/2}(Yb^{3+}) + 4I_{11/2}(Er^{3+})$ (2)

Energy transfer 2: $2F_{7/2}(Yb^{3+}) + 4G_{11/2}(Er^{3+}) \rightarrow 2F_{5/2}(Yb^{3+}) + 4F_{9/2}(Er^{3+})$ (3)

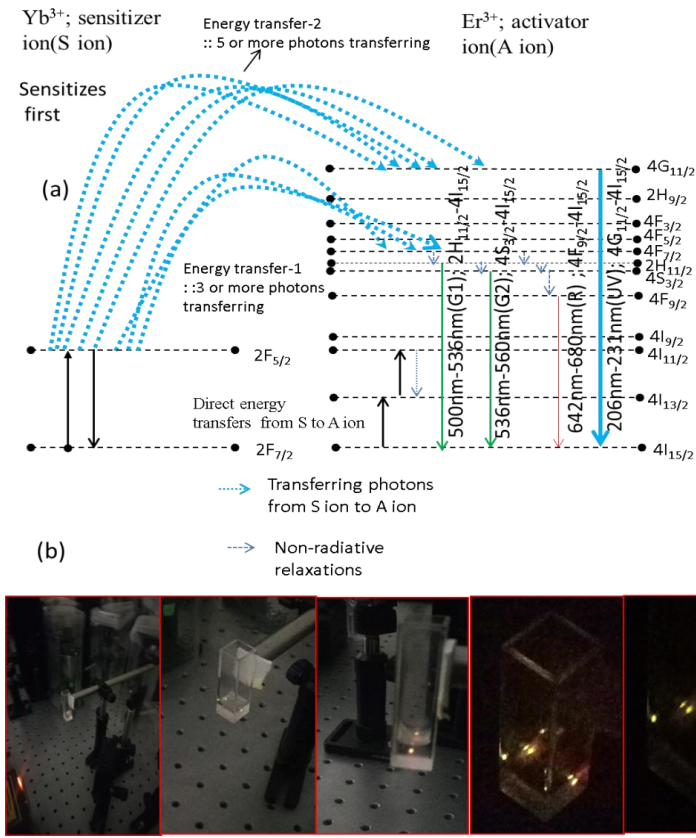

Figure 8. (a) Show energy band diagram (energy transitions) under Femtosecond laser source and (b) visible emissions in sample-cuvette during experiment (images are taken with five observations).

Further, the number of absorbed photons for red emission and for UV-light emission n(R) and n(UV) have also been calculated and found to be 5.46 and 0.40., respectively (Figure 9 (a-d)). Highest value of "n" for UV (5.46) satisfies the highest number of transferred excited photons compared to the other emissions.

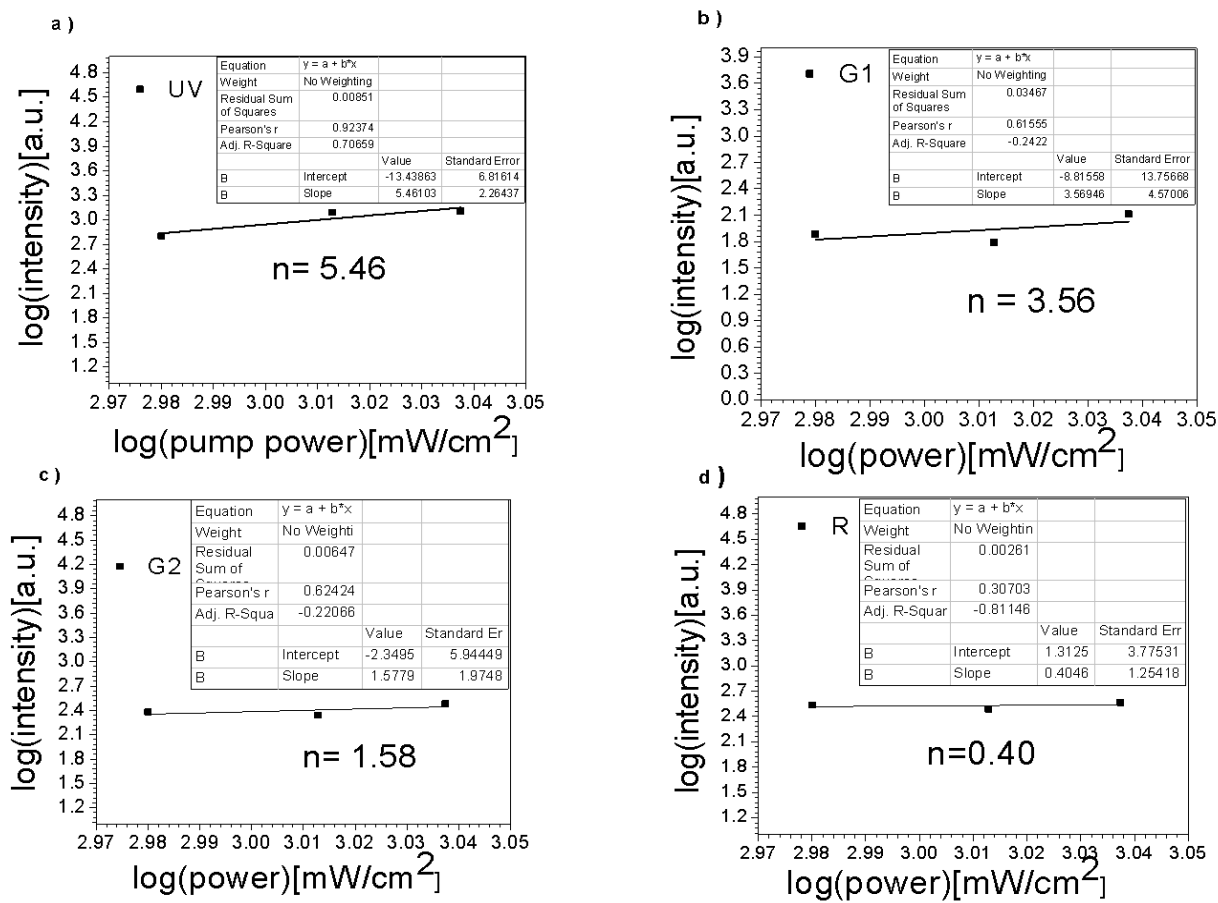

Figure 9. Dependence of Up-conversion luminescence intensity on Fs-Laser power for (a) deep-UV , (b) Green G1(c ) Green G2 and (d) Red (R) related emission (excited Fs-laser), respectively. The number of absorbed photons per photon emitted under the Fs-laser excitation power "n" value determined for the deep-UV, G1,G2 and R emissions for UCN-dot and their values also mentioned in the respective plots

The relative energy transfers for UCN-Dots in UV-Visible region occurred due to the high power Fs laser and their significant relevance along with their excitation path ways can be explained in the subsequent section.

By Fs-laser excitation sources (980, 970, 960 and 950 nm), initially the $Yb^{3+}$ sensitizes and excites to $2F_{5/2}$ level and transfers energy to $Er^{3+}$ in order to excite them to different energy levels e.g., $4I_{13/2}$, $4I_{11/2}$, $4F_{9/2}$, $4F_{7/2}$, $4G_{11/2}$ (Figure 8a). Population of $4I_{11/2}$ energy-level occurs by

19 | P a g e

transferred energy from $2F_{5/2}$ ($Yb^{3+}$) level. There are two ways to populate $4I_{13/2}$ ($Er^{3+}$) level; notably one is absorbing a 980 nm/or 970nm/or 960nm/or 950nm photons from $4I_{15/2}$ ($Er^{3+}$) energy state; another one is by non-radiative relaxation ($4I_{11/2} \rightarrow 4I_{13/2}$) between $4I_{11/2}$ and $4I_{13/2}$ energy levels. A direct energy transfer (designated as energy transfer 1) can be populated to $4F_{7/2}$ level from which non-radiative relaxations occurred to $2H_{11/2}$ level and further to $4S_{3/2}$ level; both of which are responsible for green emissions (G1 and G2, see Figure 8a) following the path: $2F_{7/2}(Yb^{3+}) \rightarrow 2F_{5/2}(Yb^{3+}) \rightarrow 4F_{7/2}(Er^{3+}) \rightarrow 2H_{11/2}/4S_{3/2}(Er^{3+}) \rightarrow 4I_{5/2}$ ($Er^{3+}$) which includes one and/or two non-radiative relaxations such as $4F_{7/2}(Er^{3+}) \rightarrow 2H_{11/2}$ and/or $2H_{11/2} \rightarrow 4S_{3/2}(Er^{3+})$. The corresponding energy transitions for G1 and G2 emissions bands are $2H_{11/2}$-$4I_{15/2}$ and $4S_{3/2}$-$4I_{15/2}$, respectively. The second direct energy transfer (energy transfer 2) occurs from $2F_{5/2}(Yb^{3+}) \rightarrow 4G_{11/2}(Er^{3+})$ level by continuous high power input excitation laser-source. From the higher excited state ($4G_{11/2}$), $Er^{3+}$ ion can be relaxed directly to the ground state ($4I_{15/2}$) with emitting high intense ultraviolet (UV) emissions corresponding to the transition $4G_{11/2} \rightarrow 4I_{15/2}$ without experiencing of non-radiative relaxations as it was observed in FL spectrum (Figure 5b).

The moderate red emission (R) assigning to the $4F_{9/2} \rightarrow 4I_{15/2}$ transition appears following the path $2F_{7/2}(Yb^{3+}) \rightarrow 2F_{5/2}(Yb^{3+}) \rightarrow 4F_{7/2}(Er^{3+}) \rightarrow 2H_{11/2} \rightarrow 4S_{3/2}(Er^{3+})$-$4F_{9/2}(Er^{3+}) \rightarrow 4I_{15/2}$ ($Er^{3+}$) including three non-radiative relaxations such as $4F_{7/2}(Er^{3+}) \rightarrow 2H_{11/2}$, $2H_{11/2} \rightarrow 4S_{3/2}(Er^{3+})$ and $4S_{3/2}(Er^{3+}) \rightarrow 4F_{9/2}(Er^{3+})$]. Thus, there are total two direct energy transfers occurred and they become responsible for UV and visible G1, G2 and R emissions which can be written in terms of the following two equations (equation 2 and 3) considering the match of energy separation.[43]

It can be noted that there is no significant change in band positions for G1, G2 or R observed with change in the Fs-laser excitation wave lengths (950 to 980 nm) (Figure 7), however change in intensity is observed (see Figure S4). However, with change in the excitation wave lengths a noticeable change in band intensities and their relative ratio for (i) peak ratio (G1/G2) and (G1/R) for emission obtained from different excitations (Figure 10a), (ii) Peak ratio(R/G2) and (R/G1)for emission obtained from different excitations (Figure 10b), (iii) peak ratio ($G_{full}$/R) or G1/R for emission obtained from different excitations (Figure 10c) and (iv) peak ratio for (UV/G1), (UV/G2) and(UV/R) for emission obtained from different excitations (Figure 10d) have been observed (see Figure S4 also).

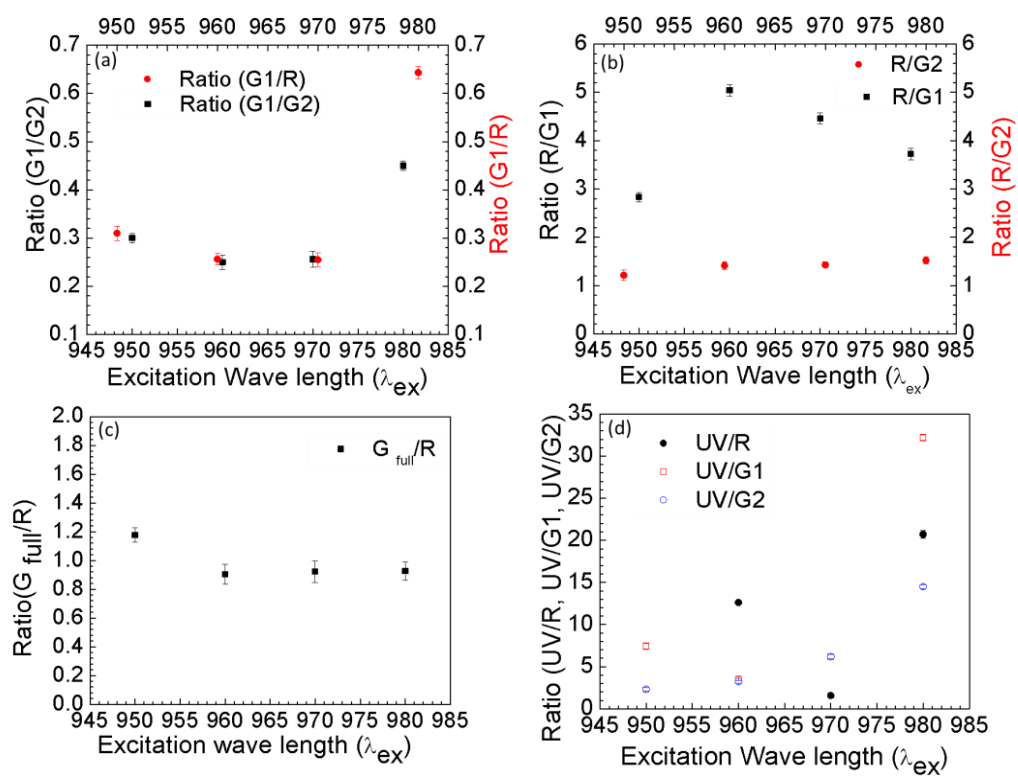

Figure 10. (a-d) Plots for peak ratios vs excitation wavelengths (nm) (a) G1/R, G1/G2 vs $\lambda_{ex}$ (b) R/G1, R/G2 vs $\lambda_{ex}$ (c) $G_{full}$ vs $\lambda_{ex}$ (d) UV/G1, UV/G2, UV/R vs $\lambda_{ex}$.

These phenomena have been observed due to the different efficient population of the energy bands responsible for the corresponding emission and due to the effective number of excited photons, i.e., n(UV), n(G1), n(G2) and n(R) (see Figure 8a, 9). The number of absorbed photons for red emission n(R) is fraction (actual 0.40) signifies that the red emission occurred only due to the cross- relaxation of photons from the higher energy state($4F_{7/2}$) to the lower energy state ($4F_{9/2}$) and not due to any other direct energy transfer.[1,35]

Appearance of strong emission bands in UV region can be explained on the basis of the change in the band-energy gap and its separation due to the interaction with Fs-laser. The band gap energy for different UV- bands due to Fs-laser interaction for the excitation wavelengths 950, 960, 970 and 980 nm have been calculated from the emission bands obtained in the Fs-luminescence spectra using the method reported elsewhere [44,45] and found to be 5.37, 5.58, 5.83 and 6.02 eV, respectively ( Figure 11 for $E_g$ calculation).

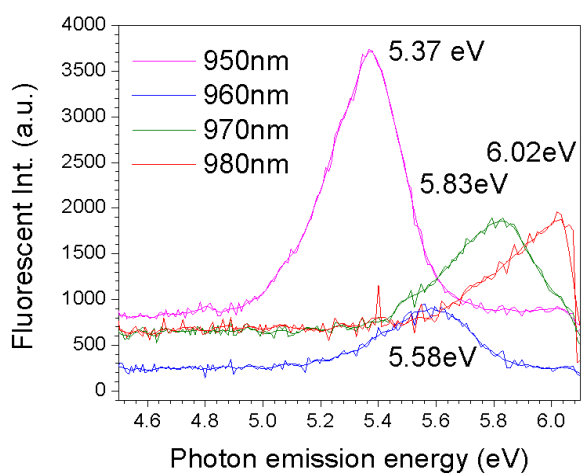

Figure 11. Band gap energies received from different fluorescent spectra under different excitation fs-laser sources (950 nm, 960 nm, 970 nm and 980 nm), shows red shifting in energy values with increasing excitation wavelengths. Fitting procedure used to obtained two-photon excitation procedure for UCN-dot at different excitation

Band gap energies received from the luminescence spectra under different excitation of Fs-laser sources, show red shifting in energy values with decreasing excitation wave lengths. It can be noted that the fitting procedure has been done with highest band energy values which corresponds to the band gap energy ($E_g$). Therefore, the shifting of bands in UV region with change in Fs-laser excitations is obvious. Further, the excitation photon energy corresponds to the Fs-laser excitation wave lengths 950nm and 960nm 970nm and 980nm have been calculated[44] and found to be 1.30, 1.29, 1.27 and 1.26 eV for which the resulted band gap energy values ($E_g$) obtained to be 5.37, 5.58, 5.83 and 6.02 eV, respectively.

Detailed result is also provided for excitation wavelengths by Fs-laser and their corresponding excitation photon energy and energy gap values in tabulated form Table S2. The band gap energy values from PL and FL spectra for a particular excitation ($\lambda = 450$ nm) wavelength has also been calculated and represented in the table (Table no. S2) and found to be in the similar order of magnitude with that obtained from the other method.

However, advent of ultrafast (UF) lasers has revolutionized the field nonlinear optical in-centro-symmetric and non-centro-symmetric materials. Electro-optic (E-O) and nonlinear phenomena are realized in the materials under the influence of strong electric field and popularly known as Pockels and Kerr effects.[46] Nonlinear phenomena are observed due to the introduction of birefringence by external E-M field and attributed to change in refractive index. The generation of high power energy upconversion from UCN-dots occurred either from the surface or interfaces of the particles. The generation of high-power upconversion emission completes in three steps, namely, (i) first incident beam of intense UF radiation induced change in the birefringence or refractive index in the materials, then (ii) the UCN-dot behave like a nonlinear Hertzian oscillator and generates polarized waves from the surface or interfaces and finally (iii) the generated polarized waves interact with each other. The interference of polarized waves generates the high energy upconversion from UCN-dots.

## CONCLUSIONS

NIR to visible upconversion fluorescent UCN-dots with cubic phase $NaYF_4$: $Yb^{+3}$; $Er^{+3}$ were synthesized by solvo-thermal decomposition of lanthanide hexahydrate precursors with organic solvents following a novel and straight forward synthesis approach. The average particle size obtained to be 3.4 nm, in dia. which is the smallest in size ever reported. These dots are stable at least for one year. UCN particles exhibit efficient upconversion fluorescence. UCN-dots are crystalline with cubic phase and exhibited visible upconversion fluorescence upon 980 nm continuous laser-excitation source. This work further revealed that UCN-dot emits both UV luminescences under interaction with Femtosecond laser. It is well known that the commercialization possibilities of UCNPs can be increased as the resulted nanocrystals are excited by 980nm laser. Though, owing to their very small sizes, the effective fluorescence intensity was reduced as a large number of cross relaxations were induced during energy-transfer mechanism but still they showed very special property while interacting with Femtosecond laser. Importantly, this work revealed that the UCN-dots could be a potential candidate for emitting highly efficient ultraviolet radiation which could be useful as UV-emitting nanophosphors and can be used in several bio-applications.[47] Since various bio-applications of UCNPs have already been reported, therefore this work directed us that the unique sizes (3.4 nm) of the UCN-dots could further open a domain of new possibilities in biomedical engineering and medical-fields for improving the treatments of infected cancer cell and tumour cells along with improving the efficiency of the device where the non-linear optical activities played an important role.

## ASSOCIATED CONTENTS

**Supporting information**

Particle size distribution, zeta potential values, EDAX, Plots of intensity [a.u.] vs excitation wavelength [nm] for UV, G1,G2 and R; Table S1(elemental analysis from EDAX), Table S2 (Band Gap Energy ($E_g$) of UCN-dots) and Calculation of the number of unit cells in UCN-dot available. This material is available free of charge via the internet at http://pubs.acs.org

Author Information

**Corresponding author**

*E-mail:paik.bme@iitbhu.ac.in, pradip.paik@gmail.com , Phone no. +91 8500109932

**Present Addresses**

ACKNOWLEDGMENT

Authors acknowledge the financial support awarded to P. Paik by DST-Nanomission, India (Ref: SR/NM/NS-1005/2015), Science and Engineering Research Board, India (Ref: EEQ/2016/000040) and M. D. Modak acknowledges the DST-INSPIRE fellowship.

REFERENCES

(1) Boyer, J.-C.; A Cuccia, L.; Capobianco, J. Synthesis of Colloidal Upconverting NaYF$_4$ : Er$^{3+}$/Yb$^{3+}$ and Tm$^{3+}$/Yb$^{3+}$ Monodisperse Nanocrystals. *Nano Lett.* **2007**, *7*, 847–852.

(2) Boyer, J.-C.; Vetrone, F.; Cuccia, L. A.; Capobianco, J. A. Synthesis of Colloidal Upconverting NaYF4 Nanocrystals Doped with Er$^{3+}$, Yb$^{3+}$ and Tm$^{3+}$, Yb$^{3+}$ via Thermal Decomposition of Lanthanide Trifluoroacetate Precursors. *J. Am. Chem. Soc.* **2006**, *128* (23), 7444–7445.

(3) Yi, G. S. and C.; M., G. Synthesis of Hexagonal-Phase NaYF4: Yb,Er and NaYF$_4$:Yb,Tm Nanocrystals with Efficient Up-Conversion Fluorescence. *Adv. Funct. Mater.* **2006**, *16*, 2324–2329.

(4) Suyver, J. F.; Grimm, J.; Krämer, K. W.; Güdel, H. U. Highly Efficient Near-Infrared to Visible up-Conversion Process in NaYF4:Er3+,Yb3+. *J. Lumin.* **2005**, *114* (1), 53–59.

(5) Li, Z.; Zhang, Y. An Efficient and User-Friendly Method for the Synthesis of Hexagonal-Phase {NaYF$_4$}:Yb, Er/Tm Nanocrystals with Controllable Shape and Upconversion Fluorescence. *Nanotechnology* **2008**, *19* (34), 345606.

(6) Salata, O. V. Applications of Nanoparticles in Biology and Medicine. *J. Nanobiotechnology* **2004**, *2* (1), 3.

(7) Patton, J. S.; Byron, P. R. Inhaling Medicines: Delivering Drugs to the Body through the Lungs. *Nat. Rev. Drug Discov.* **2007**, *6*, 67–74.

(8) Tan, W. B.; Huang, N.; Zhang, Y. Ultrafine Biocompatible Chitosan Nanoparticles Encapsulating Multi-Coloured Quantum Dots for Bioapplications. *J. Colloid Interface Sci.* **2007**, *310* (2), 464–470.

(9) Yang, S.-T.; Cao, L.; Luo, P. G.; Lu, F.; Wang, X.; Wang, H.; Meziani, M. J.; Liu, Y.; Qi, G.; Sun, Y.-P. Carbon Dots for Optical Imaging in Vivo. *J. Am. Chem. Soc.* **2009**, *131* (32), 11308–11309.

(10) Yi, D. K.; Selvan, S. T.; Lee, S. S.; Papaefthymiou, G. C.; Kundaliya, D.; Ying, J. Y. Silica-Coated Nanocomposites of Magnetic Nanoparticles and Quantum Dots Scheme 1. Synthesis of SiO2/MP-QD Nanocomposites. *J. Am. Chem. Soc* **2005**, *127*, 22.

(11) Li, K.; Qin, W.; Ding, D.; Tomczak, N.; Geng, J.; Liu, R.; Liu, J.; Zhang, X.; Liu, H.; Liu, B.; Tang, B. Z. Photostable Fluorescent Organic Dots with Aggregation-Induced Emission (AIE Dots) for Noninvasive Long-Term Cell Tracing. *Sci. Rep.* **2013**, *3*, 1150.

(12) K. Li; B. Z. Tang; B. Liu. Quantum. In *Medical Imaging: Biomedical*; 2014; Vol. 903819.

(13) Zhang, B.; Cheng, J.; Li, D.; Liu, X.; Ma, G.; Chang, J. A Novel Method to Make Hydrophilic Quantum Dots and Its Application on Biodetection. *Mater. Sci. Eng. B* **2008**, *149* (1), 87–92.

(14) Yang, Y.; Shao, Q.; Deng, R.; Wang, C.; Teng, X.; Cheng, K.; Cheng, Z.; Huang, L.; Liu, Z.; Liu, X.; Xing, B. In Vitro and In Vivo Uncaging and Bioluminescence Imaging by Using Photocaged Upconversion Nanoparticles. *Angew. Chemie Int. Ed.* **2012**, *51* (13), 3125–3129.

(15) Wang, F.; Banerjee, D.; Liu, Y.; Chen, X.; Liu, X. Upconversion Nanoparticles in Biological Labeling{,} Imaging{,} and Therapy. *Analyst* **2010**, *135* (8), 1839–1854.

(16) Cheng, L.; Wang, C.; Liu, Z. Upconversion Nanoparticles and Their Composite Nanostructures for Biomedical Imaging and Cancer Therapy. *Nanoscale* **2013**, *5* (1), 23–37.

(17) Park, Y. Il; Kim, J. H.; Lee, K. T.; Jeon, K.-S.; Na, H. Bin; Yu, J. H.; Kim, H. M.; Lee, N.; Choi, S. H.; Baik, S.-I.; Kim, H.; Park, S. P.; Park, B.-J.; Kim, Y. W.; Lee, S. H.; Yoon, S.-Y.; Song, I. C.; Moon, W. K.; Suh, Y. D.; Hyeon, T. Nonblinking and Nonbleaching Upconverting Nanoparticles as an Optical Imaging Nanoprobe and T1 Magnetic Resonance Imaging Contrast Agent. *Adv. Mater.* **2009**, *21* (44), 4467–4471.

(18) Wang, H.-Q.; Batentschuk, M.; Osvet, A.; Pinna, L.; Brabec, C. J. Rare-Earth Ion Doped Up-Conversion Materials for Photovoltaic Applications. *Adv. Mater.* **2011**, *23* (22- 23), 2675–2680.

(19) Atwater, H. A.; Polman, A. Plasmonics for Improved Photovoltaic Devices. *Nat. Mater.* **2010**, *9*, 205–213.

(20) Schietinger, S.; Aichele, T.; Wang, H.-Q.; Nann, T.; Benson, O. Plasmon-Enhanced Upconversion in Single NaYF$_4$:Yb$^{3+}$/Er$^{3+}$ Codoped Nanocrystals. *Nano Lett.* **2010**, *10* (1), 134–138.

(21) Atre, A. C.; Garcia-Etxarri, A.; Alaeian, H.; Dionne, J. A. Toward High-Efficiency Solar Upconversion with Plasmonic Nanostructures. *J. Opt.* **2012**, *14* (2), 24008.

(22) Eichelbaum, M.; Rademann, K. Plasmonic Enhancement or Energy Transfer? On the Luminescence of Gold-, Silver-, and Lanthanide-Doped Silicate Glasses and Its Potential for Light-Emitting Devices. *Adv. Funct. Mater.* **2009**, *19* (13), 2045–2052.

(23) Huang, X. Broadband Dye-Sensitized Upconversion: A Promising New Platform for Future Solar Upconverter Design. *J. Alloys Compd.* **2017**, *690*, 356–359.

(24) Schäfer, H.; Ptacek, P.; Eickmeier, H.; Haase, M. Synthesis of Hexagonal $Yb^{3+}$,$Er^{3+}$-Doped $NaYF_4$ Nanocrystals at Low Temperature. *Adv. Funct. Mater.* **2009**, *19* (19), 3091–3097.

(25) Hall, B. D.; Zanchet, D.; Ugarte, D. Estimating Nanoparticle Size from Diffraction Measurements. *J. Appl. Crystallogr.* **2000**, *33* (6), 1335–1341.

(26) Mai, H.-X.; Zhang, Y.-W.; Sun, L.-D.; Yan, C.-H. Highly Efficient Multicolor Up-Conversion Emissions and Their Mechanisms of Monodisperse $NaYF_4$:Yb,Er Core and Core/Shell-Structured Nanocrystals. *J. Phys. Chem. C* **2007**, *111* (37), 13721–13729.

(27) MacKenzie, L. E.; Goode, J. A.; Vakurov, A.; Nampi, P. P.; Saha, S.; Jose, G.; Millner, P. A. The Theoretical Molecular Weight of $NaYF_4$:RE Upconversion Nanoparticles. *bioRxiv* **2017**, 114744.

(28) Klier, D. T.; Kumke, M. U. Analysing the Effect of the Crystal Structure on Upconversion Luminescence in $Yb^{3+}$,$Er^{3+}$-Co-Doped $NaYF_4$ Nanomaterials. *J. Mater. Chem. C* **2015**, *3* (42), 11228–11238.

(29) Structural Effects on Optical Properties of Fcc $NaYF_4$:Yb,Er Nanoparticles; *Chapter-4;* *https://scholarbank.nus.edu.sg/bitstream/* 10635/34459/5/04chap.pdftle.

(30) Wilhelm, S.; Hirsch, T.; Patterson, W. M.; Scheucher, E.; Mayr, T.; Wolfbeis, O. S. Multicolor Upconversion Nanoparticles for Protein Conjugation. *Theranostics* **2013**, *3* (4), 239–248.

(31) Assaaoudi, H.; Shan, G.-B.; Dyck, N.; Demopoulos, G. P. Annealing-Induced Ultra-Efficient NIR-to-VIS Upconversion of Nano-/Micro-Scale α and β NaYF$_4$:Er$^{3+}$,Yb$^{3+}$ Crystals. *CrystEngComm* **2013**, *15* (23), 4739–4746.

(32) Nyk, M.; Kumar, R.; Ohulchanskyy, T. Y.; Bergey, E. J.; Prasad, P. N. High Contrast in Vitro and in Vivo Photoluminescence Bioimaging Using Near Infrared to Near Infrared Up-Conversion in Tm$^{3+}$ and Yb$^{3+}$ Doped Fluoride Nanophosphors. *Nano Lett.* **2008**, *8* (11), 3834–3838.

(33) Chen, G.; Ohulchanskyy, T. Y.; Kumar, R.; Ågren, H.; Prasad, P. N. Ultrasmall Monodisperse NaYF$_4$:Yb$^{3+}$/Tm$^{3+}$ Nanocrystals with Enhanced Near-Infrared to Near-Infrared Upconversion Photoluminescence. *ACS Nano* **2010**, *4* (6), 3163–3168.

(34) Lee, M.; Park, Y. H.; Kang, E. B.; Chae, A.; Choi, Y.; Jo, S.; Kim, Y. J.; Park, S.-J.; Min, B.; An, T. K.; Lee, J.; In, S.-I.; Kim, S. Y.; Park, S. Y.; In, I. Highly Efficient Visible Blue-Emitting Black Phosphorus Quantum Dot: Mussel-Inspired Surface Functionalization for Bioapplications. *ACS Omega* **2017**, *2* (10), 7096–7105.

(35) Schietinger, S.; Menezes, L. de S.; Lauritzen, B.; Benson, O. Observation of Size Dependence in Multicolor Upconversion in Single Yb$^{3+}$, Er$^{3+}$ Codoped NaYF$_4$ Nanocrystals. *Nano Lett.* **2009**, *9* (6), 2477–2481.

(36) Wang, F.; Deng, R.; Wang, J.; Wang, Q.; Han, Y.; Zhu, H.; Chen, X.; Liu, X. Tuning Upconversion through Energy Migration in Core–shell Nanoparticles. *Nat. Mater.* **2011**, *10*, 968–973.

(37) Aktsipetrov, O. A.; Dolgova, T. V; Fedyanin, A. A.; Murzina, T. V; Inoue, M.; Nishimura, K.; Uchida, H. Magnetization-Induced Second- and Third-Harmonic Generation in Magnetophotonic Crystals. *J. Opt. Soc. Am. B* **2005**, *22* (1), 176–186.

(38) Yu, W.; Xu, W.; Song, H.; Zhang, S. Temperature-Dependent Upconversion Luminescence and Dynamics of NaYF$_4$:Yb$^{3+}$/Er$^{3+}$ Nanocrystals: Influence of Particle Size and Crystalline Phase. *Dalt. Trans.* **2014**, *43* (16), 6139–6147.

(39) Shan, J.; Uddi, M.; Wei, R.; Yao, N.; Ju, Y. The Hidden Effects of Particle Shape and Criteria for Evaluating the Upconversion Luminescence of the Lanthanide Doped Nanophosphors. *J. Phys. Chem. C* **2010**, *114* (6), 2452–2461.

(40) Lu, D.; Cho, S. K.; Ahn, S.; Brun, L.; Summers, C. J.; Park, W. Plasmon Enhancement Mechanism for the Upconversion Processes in NaYF$_4$:Yb$^{3+}$,Er$^{3+}$ Nanoparticles: Maxwell versus Förster. *ACS Nano* **2014**, *8* (8), 7780–7792.

(41) J.F.Suyver, A. Aebischer, S. Garcia-Revilla, P. G. and H. U. G. 2005, 71,. *Phy. Rev. B. Condens. Matter Phys.* **2005**, *71*, 1–9.

(42) Kingsley, J. D.; Fenner, G. E.; Galginaitis, S. V. Kinetics and efficiency of infrared-to-visible conversion in LaF$_3$:Yb,Er. *Appl. Phys. Lett.* **1969**, *15* (4), 115–117.

(43) Song, H.; Sun, B.; Wang, T.; Lu, S.; Yang, L.; Chen, B.; Wang, X.; Kong, X. Three-Photon Upconversion Luminescence Phenomenon for the Green Levels in Er$^{3+}$/Yb$^{3+}$ Codoped Cubic Nanocrystalline Yttria. *Solid State Commun.* **2004**, *132* (6), 409–413.

(44) Li, L.; Hu, J.; Yang, W.; Alivisatos, A. P. Band Gap Variation of Size- and Shape-Controlled Colloidal CdSe Quantum Rods. *Nano Lett.* **2001**, *1* (7), 349–351.

(45) Dukovic, G.; Wang, F.; Song, D.; Sfeir, M. Y.; Heinz, T. F.; Brus, L. E. Structural Dependence of Excitonic Optical Transitions and Band-Gap Energies in Carbon Nanotubes. *Nano Lett.* **2005**, *5* (11), 2314–2318.

Please do not adjust margins(46) Melnichuk, M.; Wood, L. T. Direct Kerr Electro-Optic Effect in Noncentrosymmetric Materials. *Phys. Rev. A* **2010**, *82* (1), 13821.

(47) Chien, Y.-H.; Chan, K. K.; Yap, S. H. K.; Yong, K.-T. NIR-Responsive Nanomaterials and Their Applications; Upconversion Nanoparticles and Carbon Dots: A Perspective. *J. Chem. Technol. Biotechnol.* **2018**, *93* (6), 1519–1528.
31 | P a g e

Please do not adjust margins

## TABLE OF CONTENTS (TOC Figure)

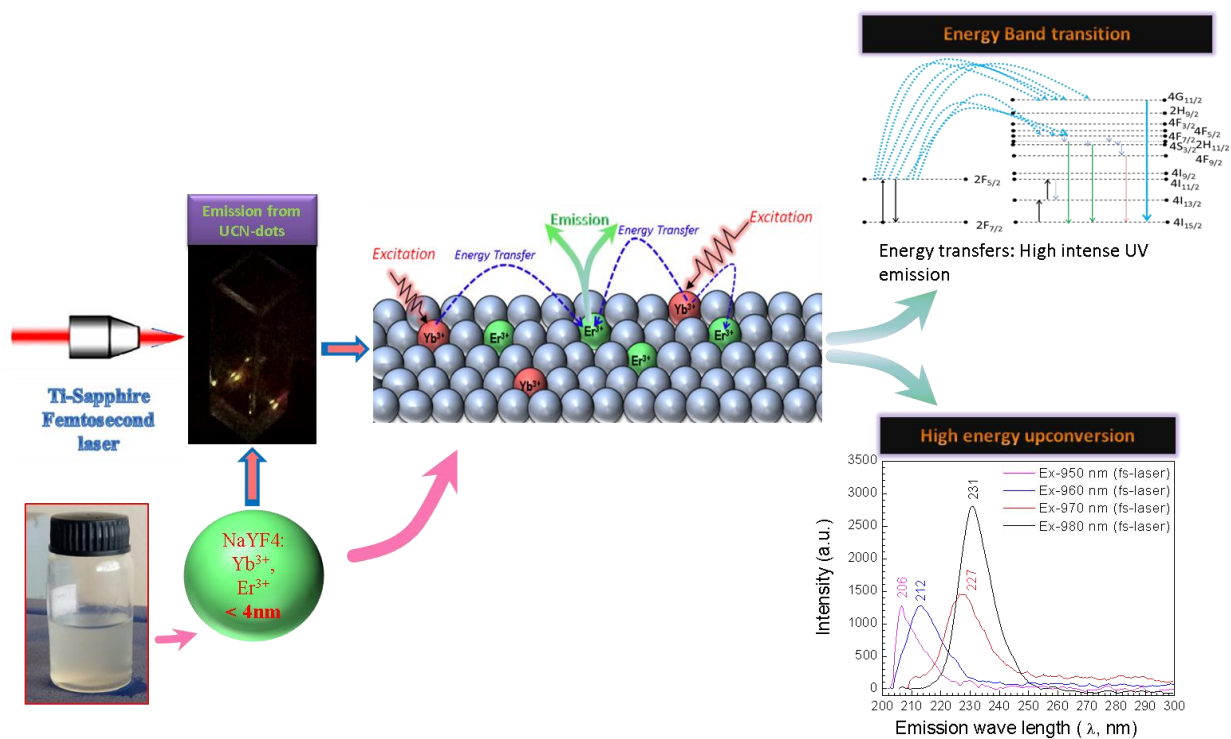

**TOC Text**: In nanodots higher energy upconversion is occurred under Femtosecond laser source

# Upconverting nanodots of NaYF$_4$:Yb$^{3+}$Er$^{3+}$; Synthesis, characterization and UV-Visible luminescence study through Ti: sapphire 140-femtosecond laser–pulses

*Monami Das Modak[a], Ganesh Damarla[b], K Santhosh Kumar[a], Somedutta Maity[a], Anil K Chaudhury[b] and Pradip Paik[a,c]\**

[a.]School of Engineering Sciences and Technology, University of Hyderabad, Hyderabad, Telangana, PIN: 500046

[b.] Advanced Center of Research in High Energy Materials, University of Hyderabad, Hyderabad, Telangana, PIN: 500046 Address here

[c.]School of Biomedical Engineering, Indian Institute of Technology (IIT)-BHU, Varanasi, UP, PIN 221005

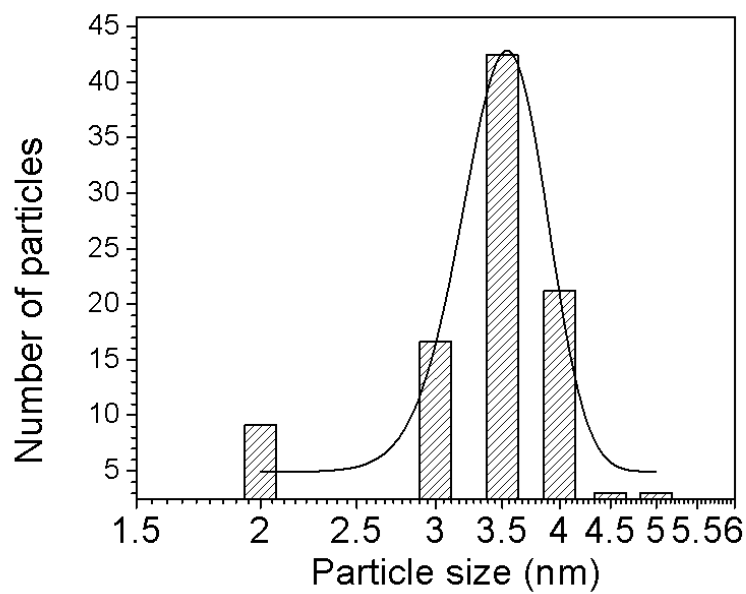

**Figure S1.** Histogram particle size and distribution calculated from TEM micrograph

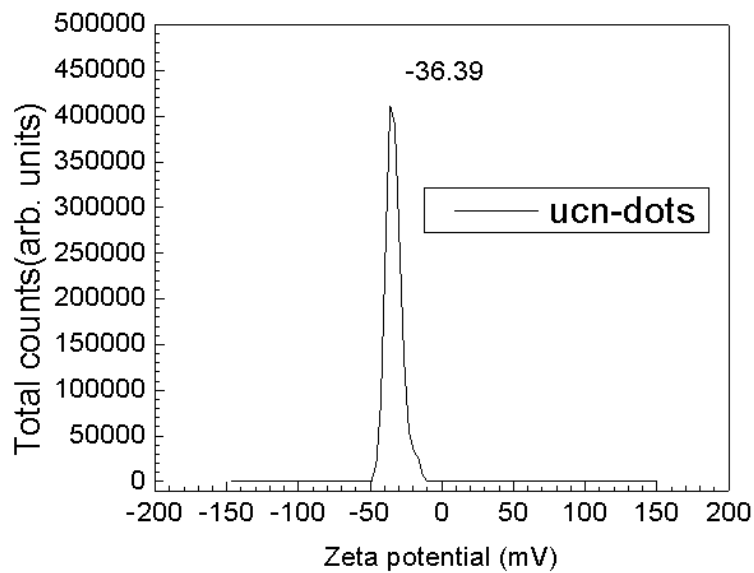

**Figure S2.** Zeta potential for colloidal UCN-dot sample

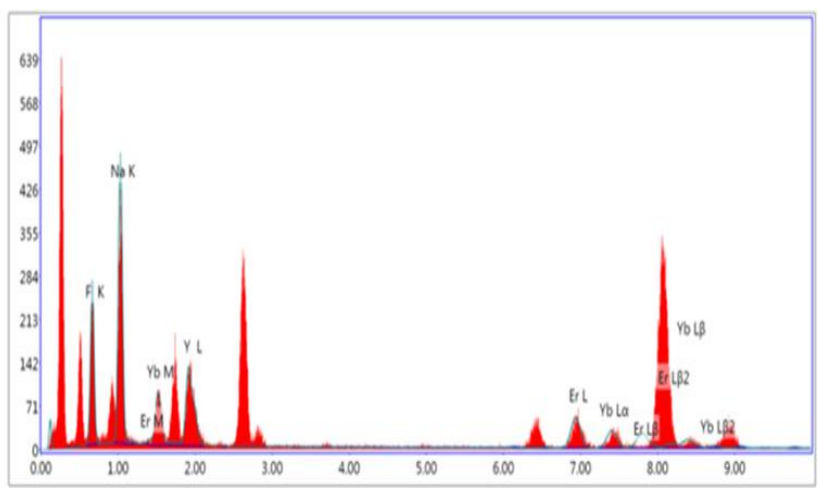

**Figure S3.** EDAX analysis of UCN-dots colloidal sample

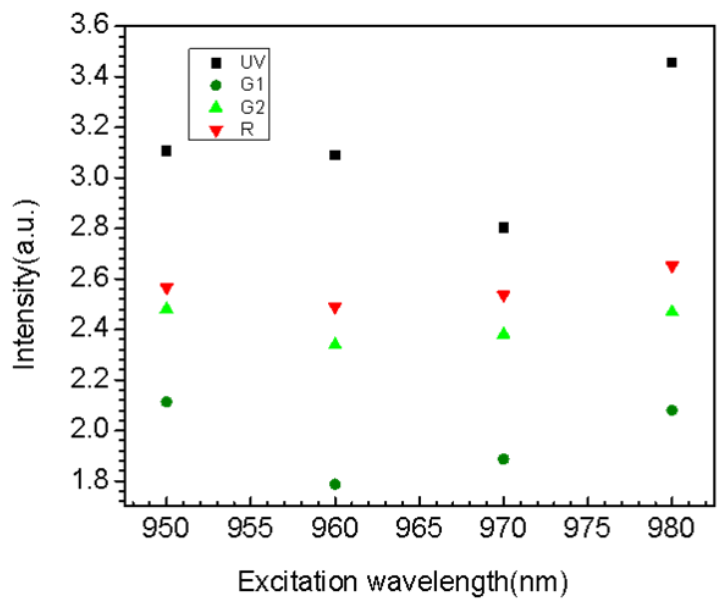

**Figure S4.** Plots intensity [a.u.] vs excitation wavelength[nm] for UV,G1,G2,R

| Element | Weight(%) | Atomic (%) |
|---|---|---|
| F | 25.23 | 29.33 |
| Na | 73.34 | 70.47 |
| Er | 0.85 | 0.11 |
| Yb | 0.43 | 0.05 |
| Y | 0.16 | 0.04 |

**Table S1.** Table for elemental analysis from EDAX

| Types of Spectra | Excitation Wavelength $\lambda_{ex}$(nm) | Excited photon energy (eV) [1240/Wavelength] | Band gap values(eV) |
|---|---|---|---|
| Femtosecond Laser(Fs)-spectra | 950nm | 1.31 eV | 5.37 eV |
| | 960nm | 1.29eV | 5.58eV |
| | 970nm | 1.28eV | 5.83eV |
| | 980m | 1.27eV | 6.02eV |
| Fluorescence-spectra(FL) | 980nm | 1.27eV | 2.5eV, 2.35eV,2.28eV, 1.87eV,1.48eV |
| Photoluminescence spectra(PL) | 450nm | 2.76eV | 3.26eV, 3.05eV,2.88eV, 2.68eV, 2.38eV, 2.26eV,2.18eV, 2.04eV 1.94eV |

**Table S2.** Band Gap Energy (Eg) of UCN-dots with different excitation photon energies (Fs) and single photon energy (FL,PL)

The number of unit cells in UCN-dot:

The average lattice constant has been found to be = 5.39 Å

Assuming UCN-dots to be almost spherical, the volume of one particles has been calculated as= $V_{UCN\text{-}dots} = 4/3\pi r^3 = 4/3 \times 3.14 \times (1.75)^3 = 22.4379\ nm^3$ [ Considering average particle diameter as 3.5nm, radius= 3.5/2= 1.75nm]

Now, As the crystal structure is cubic-FCC, UCN-dots consist of cubic unit cells, so the volume of the cubic unit cell= $V_{unit\ cell} = (a_o)^3 = (5.39 Å)^3 = 156.590 \times 10^{-3}\ nm^3$.

Therefore, the number of unit cells in one UCN-dot particle= $V_{UCN\text{-}dots}/V_{unit\ cell}$ = 0.143 ×10³ ~ almost 143 no. of unit cells.

[Note: In cubic cell, Number of F⁻ ions  = 8, so in one particle 1144 no. of F⁻ions are present

Number of Na⁺/RE³⁺ ions  = randomly occupy the positions

According to previous study, please follow the link http://dx.doi.org/10.1101/114744

As standardized calculation, with rare earth dopants ($Yb^{3+}$, $Er^{3+}$), number of effective Na, F, Yb, Er atoms present are= 2, 8, 1, 1 as fraction of $Y^{3+}$ ions are substituted by rare earth ions($Yb^{3+}$ and $Er^{3+}$ ) [ref-as mentioned above] .